\documentclass[12pt,preprint]{aastex}
\begin{document}

\title{On the Detection of High Redshift Black Holes with ALMA through CO and H$_2$ Emission}

  \author{Marco Spaans} \affil{Kapteyn Astronomical Institute,
  P.O. Box 800, 9700 AV Groningen, The Netherlands}
  \email{spaans@astro.rug.nl}

\and

  \author{Rowin Meijerink} \affil{Astronomy Department, University of
  California, Berkeley, CA 94720, United
  States}

\begin{abstract}

Many present-day galaxies are known to harbor supermassive,
$\ge 10^6$ $M_\odot$, black holes.
These central black holes must have grown through accretion from less massive
seeds in the early universe. The molecules CO and H$_2$ can be used
to trace this young population of accreting massive black holes through the
X-ray irradiation of ambient gas.
The X-rays drive a low-metallicity ion-molecule chemistry that leads
to the formation and excitation of CO and H$_2$ in $100<T\le 1,000$ K gas.
H$_2$ traces very low metallicity gas, $\sim 10^{-3}$ solar or less, while some
pollution by metals, $\sim 10^{-2}$ solar or more, must have taken place to
form CO. Strong CO $J>15$ and H$_2$ S(0) \& S(1) emission is found that allows
one to constrain ambient conditions.
Comparable line strengths cannot be produced by FUV or cosmic ray irradiation.
Weak, but perhaps detectable, H$_3^+$ (2,2)$\rightarrow$(1,1) emission is
found and discussed. The models predict that black hole masses larger than
$10^{5}$ $M_\odot$ can be detected with ALMA, over a redshift range of 5-20,
provided that the black holes radiate close to Eddington.

\end{abstract}


\section{Introduction}

Central to the study of galaxy evolution is the formation and evolution
of the supermassive black holes in their nuclei.
Accretion onto these black holes can provide the energy source for active
galactic nuclei, which in turn impact the evolution of galaxies (Silk 2005).
The processes believed to play a role in the formation of seed black holes,
from which large black holes may form through accretion, involve
1) dynamical friction and collision processes in dense young stellar clusters
(Portegies Zwart et al.\ 2004); 2) seeds as the remnants of popIII stars
(e.g., Bromm et al.\ 1999, Abel et al.\ 2000, Yoshida et al.\ 2003);
3) accretion of low angular momentum material and gravitational instability
in primordial disks (e.g., Koushiappas, Bullock \& Dekel 2004;
Lodato \& Natarajan 2007); and 4) the (singular) collapse of massive
pre-galactic halos (Bromm \& Loeb 2003; Spaans \& Silk 2006).

The growth of these seed black holes to larger sizes involves accretion that
roughly follows an Eddington rate and requires the incorporation of feedback
effects (Silk \& Rees 1998, Wyithe \& Loeb 2003, Di Matteo, Springel \&
Hernquist 2005). See Pelupessy, Di Matteo \& Ciardi (2007) for an asessment
on the difficulties that seed black holes have to grow at the Eddingtion rate.
Understanding the growth of these black holes is also important because there
appears to be a scaling relation between bulge and black hole mass, with
about $10^{-3}$ of the bulge mass tied up in the central black holes
(Magorrian et al.\ 1998, Ferrarese \& Merritt 2000, H\"aring \& Rix 2004).
In this work, it is investigated how one can observe a population of these
putative black holes in the early universe, at redshifts $z=5-20$, through
high temperature molecular lines that are driven by X-rays and that
are accessible to the Atacama Large Millimeter Array (ALMA), which covers the
300 $\mu$m to 3 mm wavelength range.

\section{Model Description}

We consider a high redshift halo which already contains, by assumption, a
seed black hole.
Suppose further that an initially metal-free hydrogen gas is cooled
by Lyman $\alpha$ to $T\sim 10^4$ K and settles in the center of a halo
close to the black hole. Part of the gas is likely
to experience a period of popIII star formation. If popIII star formation
occurs, then the short life times of primordial stars ensure that the accreted
gas is quickly polluted by a modest amount of dust and metals.
If no star formation takes place, then the gas will remain metal-free.
In either case, the accretion process will lead to the emission of X-rays
that impact the thermal, ionization an chemical balance of the gas in the
halo, leading to an X-ray dominated region (XDR, Maloney et al.\ 1996;
Meijerink \& Spaans 2005).
For simplicity the black hole is taken to radiate at the Eddington luminosity.

For a baryonic number density
$\rho_b/m_H=3\times 10^{-7}$ cm$^{-3}$ today, hydrogen mass $m_H$,
halo masses of $M_h=10^7-10^9$ $M_\odot$ and a characteristic size scale
of $L=(3M_h/4\pi\rho)^{1/3}$, one has a typical mean
density and column of $n_0=0.05[(1+z)/10]^3$ cm$^{-3}$ and
$N_0=10^{22}[(1+z)/10]^2(M_h/10^9M_\odot)^{1/3}$ cm$^{-2}$, respectively.
The work of Mo, Mao \& White (1988) shows that the subsequent formation of a
disk occurs, with a collapse factor of $1/\lambda =0.05$. This yields
densities that exceed $10^{2.5}$ cm$^{-3}$ within 1 kpc.
The above cosmology provides the boundary conditions for the ambient density
and column density of individual models, and the metallicity is put in by hand.

The models of Meijerink \& Spaans (2005) and Meijerink et al.\ (2007)
are used to compute the thermal, chemical and ionization balance of
the irradiated gas self-consistently, for one-dimensional constant density
slabs of gas.
The multi-zone escape probability method of Poelman \& Spaans (2005,
2006) has been used to compute the line intensities presented here.
The same radiative transfer is performed in relevant atomic (fine-structure)
and molecular (rotational and vibrational) cooling lines.
Cloud type ``A'' from Meijerink et al.\ (2007, their Table 1) is adopted,
which is 1 pc in size. The interested reader is referred to the cited papers
for a detailed description of all physical processes involved.
There are four free parameters in the models: hydrogen density, hydrogen
column density, metallicity and X-ray flux. The latter parameterizes
unknowns like the accretion rate, turbulent viscosity and the spectral
shape of the X-ray radiation. For definiteness, a power law radiation field
with $E^{-0.9}$ is adopted for energies $E$ between 1 and 100 keV,
appropriate for a self-absorbed Seyfert nucleus.
This slope, if it is between $-1.1$ and $-0.7$ does not significantly
impact the chemistries of H$_2$, CO and H$_3^+$; see also Meijerink \& Spaans
(2005) for the case of a 1 keV thermal spectrum.
Solar elemental abundance ratios are
adopted. As long as [O/C]$>1$ this does not influence the CO results.
Chemical equilibrium is assumed. At densities of $\sim 10^5$ cm$^{-3}$ and for
high X-ray fluxes, collisional and radiative time scales are short compared to
the free-fall time.
The chemical network comprises a few thousand reactions between 154 species
with sizes up to 4 atoms (Woodall et al.\ 2006). Polycyclic aromatic
hydrocarbons and small grains are included in the charge balance and are
assumed to scale with the elemental carbon abundance.

Following the above cosmology, we consider column densities of
$10^{22}-10^{24}$ cm$^{-2}$ and densities of $n=10^3-10^5$ cm$^{-3}$.
The X-ray flux takes values of $F_X=0.1-100$ erg s$^{-1}$ cm$^{-2}$.
Results are shown in the Figures 1, 2 and 3 for a density of $10^5$ cm$^{-3}$.
Values below this density but above $10^3$ cm$^{-3}$ were found to lead to
similar signal strengths for H$_2$ and H$_3^+$. For CO, line intensities
smaller by a factor of $\sim n/10^5$ were found.
Recall that the Eddington luminosity is
$L_{\rm edd}\approx 1.3\times 10^{38} M/M_\odot$ erg s$^{-1}$.
So $F_X\approx 100$ erg s$^{-1}$ cm$^{-2}$ corresponds to a $10^6$
$M_\odot$ black hole that emits $\sim 10^{44}$ erg s$^{-1}$ through a surface
with a radius of about 100 pc.
The level to which popIII star formation pollutes the center of the primodial
galaxy with metals through supernova explosions is a free parameter.
Metallicities between $10^{-3}$ and 1.0 solar are considered since $10^{-3}$
solar is quite like a zero-metallicity gas as far as H$_2$ and CO are
concerned, and supersolar values
appear unlikely for the bulk of the very high redshift gas.
Dust grains, with standard Milky Way properties (Mathis et al.\ 1977), are
included and their abundance is assumed to scale with the overall metallicity.
The velocity dispersion of the gas has a thermal contribution, equal
to $\Delta V=12.9(T/10^4)^{1/2}$ km/s, and a turbulent
contribution, equal to $\Delta V=5$ km/s, for individual gas clouds on
the scale of 1 pc.
In the dense gas considered here, turbulence is expected to be maintained at
a level similar to that in active galaxies.

\section{Results}

The main goal is to compute the expected columns of H$_2$, CO and
H$_3^+$, and their associated emission strengths.
For simplicity, a fiducial column of $10^{23.5}$ cm$^{-2}$ is adopted for the
predicted line emissivities. This is driven by the theoretical considerations
above, but also by recent observations of massive galaxies at $z\sim 2$
(Daddi et al.\ 2007a,b). A significant fraction (20-30\%) of the systems
appear to contain heavily obscured AGN with columns in excess of $10^{24}$
cm$^{-2}$. These massive systems appear to be a somewhat later stage of the
concurrent bulge-black hole mass forming systems studied here. The X-ray
luminosities of these lower redshift counterparts is $(1-4)\times 10^{43}$
erg s$^{-1}$ in the 2-8 keV band. The precise value of the total obscuring
hydrogen column does not impact our emissivity results as long as it exceeds
$10^{23}$ cm$^{-2}$, i.e., includes all $T>100$ K gas.

\subsection{H$_2$ and H$_3^+$}

In the absence of dust grains, H$_2$ is formed in the gas phase through the
H$^-$ route, H$^-$+H$\rightarrow$H$_2$+e$^-$. This leads to high abundances of
H$_2$, $10^{-2}-10^{-0.5}$, even for low metallicities (Figure 1). So,
contrary to far-ultraviolet (FUV, 6-13.6 eV) illumination, X-ray irradiation
constitutes a form of positive feedback for H$_2$ (Haiman et al.\ 1997).
The models with a modest metallicity of $>10^{-3}$, and thus dust grains,
follow the H$_2$ formation prescription as in Cazaux \& Spaans (2004),
which includes both physisorbed and chemisorbed hydrogen atoms.
The columns of H$_2$ that are reached at low metallicities are as large
as $10^{23}$ cm$^{-2}$. This while temperatures exceed 100 K over the bulk of
this column, and reach $10^3$ K at its edge, sufficient to excite the
$\sim 500$ K S(0) line at 28 $\mu$m and $\sim 800$ K S(1) line at 17 $\mu$m.
The S(1) line is typically weaker than the S(0) line, while both lines are
optically thin and in LTE. High molecular gas temperatures are reached in
XDRs because of efficient ionization and Coulomb heating, rather than
photo-electric heating from dust grains (Meijerink \& Spaans 2005).
Interestingly, an increase in metallicity decreases the H$_2$ line strength.
This is a direct consequence of enhanced cooling in fine-structure
lines and a lower resulting temperature. Hence, the pure rotational H$_2$ lines
are particularly well suited to detect the earliest stages of black hole
accretion, prior to significant metal pollution by star formation.
Metallicities below $10^{-3}$ were found to lead to quite similar H$_2$
emissivities since all H$_2$ is formed in the gas phase and fine-structure
cooling is modest.

Figure 2 (see Section 4 for its details) shows rest frame line intensities of
$\sim 10^{-1}$ erg s$^{-1}$ cm$^{-2}$ sr$^{-1}$ at a metallicity of $10^{-3}$
solar.
For the numbers below, the ALMA sensitivity tool on the ESO website has been
used and the concordance model, with the latest WMAP3 results, is adopted.
With a source size of $0.2''$, or 600 pc at $z=15$, this yields a S(0) flux
density of $39/(1+z)^3$ Jy, for a fiducial galaxy center velocity dispersion of
20 km/s ($\sim 200$ pc from a $\sim 10^8 M_\odot$ central mass). For the S(0)
line from $z=15$, so at 0.45 mm, and for 10 km/s
spectral resolution, this is detectable with ALMA (50 antennas, beam
size = source size) at the $6\sigma$ level in 4 hours of integration.
Hence, despite the fact that the line falls in the less sensitive band 9,
spectrally resolved detection is possible. The pure rotational
H$_2$ lines are intrinsically very bright, because of the strong contribution
from the X-ray driven H$^-$ route, and are accessible to ALMA for redshifts
above 10, for S(0) 28 $\mu$m, and above 16 for S(1) 17 $\mu$m.

Figure 1, for a density of $10^5$ cm$^{-3}$, shows that the H$_3^+$ abundance
increases with ionization parameter $F_X/n$ for metallicities below $10^{-2}$
of solar. A higher X-ray ionization rate leads to a larger H$_3^+$ abundance,
through H$_2$ secondary ionizations, and boosts the formation rate
H$_2$+H$_2^+$$\rightarrow$H$_3^+$.
The H$_2$ abundance is thus key since no H$_3^+$ can be formed without it.
Too large values of $F_X/n$ lead to a decrease in the H$_3^+$ abundance,
driven by dissociative recombination.

The best candidate for an H$_3^+$ detection is the optically thin
(2,2)$\rightarrow$(1,1) line at 95 $\mu$m, as suggested by Pan \& Oka (1986).
The critical density of the (2,2)$\rightarrow$(1,1) transition is about
$2\times 10^3$ cm$^{-3}$ and the excitation energy $\sim 150$ K.
Collisions between H$_3^+$ and electrons have been included in the non-thermal
rotational excitation of H$_3^+$ (Faure et al.\ 2006).
In all, Figure 2 shows that one reaches a maximum rest frame intensity of
$\sim 10^{-6}$ erg s$^{-1}$ cm$^{-2}$ sr$^{-1}$, at low metallicity and
strong X-ray irradiation, with typical abundances of $10^{-8}-10^{-9}$.
With a source size of $0.4''$, or about 1.8 kpc at $z=9$, this yields a flux
density of $9/(1+z)^3$ mJy, for a fiducial galaxy center velocity dispersion
of 20 km/s. For $z=10$ at 1 mm (from 95 $\mu$m), this is barely detectable
with ALMA at the $3\sigma$ level, in 48 hours of integration (50 antennas,
beam size = source size), and only if the line is unresolved.

\subsection{CO}

Any metallicity larger than $10^{-3}$ solar leads to significant,
$\sim 10^{-8}-10^{-6}$, abundances of CO.
X-rays, because of their large energy, do not dissociate CO directly.
FUV photons are produced through collisional excitation of H and H$_2$ by
electrons, followed by radiative decay. This UV flux is generally
modest and thus CO can survive even in strong X-ray radiation fields.
Given the high molecular gas temperatures in XDRs, $10^2-10^3$ K,
rotational levels with $J>10$ are excited (Meijerink et al.\ 2006, 2007).
This very high $J$ CO emission requires densities $>10^{4.5}$ cm$^{-3}$
because the critical densities of these lines are about $10^6$ cm$^{-3}$.
Metallicities in excess of $10^{-2}$ of solar further raise the CO
emissivities, even though the higher abundances of C, O, Si and Fe also
enhance the cooling of the gas (Santoro \& Shull 2006).
The CO emissivities below $10^{-3}$ solar are negligible.

One finds from Figure 2 that, at a metallicity $>10^{-2}$ solar, rest frame
CO line intensities reach $\sim 10^{-3}$ erg cm$^{-2}$ s$^{-1}$
sr$^{-1}$, which yields a flux density of $\sim 1.4/(1+z)^3$ Jy for a $0.2''$
source size, or 850 pc at $z=10$, and a fiducial galaxy center velocity
dispersion of 20 km/s.
For the rest frame peak in the CO line spectral energy distribution (SED)
at 3000 GHz (see Figure 2), 10 km/s velocity resolution and
for $z=10$, ALMA detects such a source at $5\sigma$ in 8 hours
of integration (50 antennas, beam size = source size). Spectrally resolved
detection is possible. ALMA covers 0.3 to 3.0 mm, fortuitously matching most
of the X-ray driven CO line SED for $z=5-20$.

\section{Discussion}

Figure 2 shows how the rest frame spectral line distribution of CO, pure
rotational H$_2$ and H$_3^+$ evolves with irradiation and metallicity for an
ALMA beam that is filled with 1 pc clouds at $\sim 10^5$ cm$^{-3}$. These
clouds are in Keplerian orbit around the central black hole and are randomly
distributed over a spherical region with a linear size of 1 kpc in such a
way that there is about one cloud along each line of sight.
This central region is slowly enriched in metals as indicated.
The total column density is on average $\sim 10^{23.5}$ cm$^{-2}$ along each
line of sight.
The total intensity is then found by a ray-trace on the level populations of
the individual clouds.
More complicated geometries are not an issue as long as the cloud covering
factor is of the order of unity and the bulk of the emitted X-rays are
absorbed.

Overall, X-ray fluxes of $>1$ erg s$^{-1}$ cm$^{-2}$, corresponding
to black hole masses of $>10^{5}$ $M_\odot$ and a region with a size of $>600$
pc yield detectable lines, provided the black hole is radiating at its
Eddington luminosity. This mass value is comparable to that of black holes
present in in cosmological simulations of $10^8-10^9M_\odot$ halos
collapsing at $10<z<20$ (Peluppessy et al.\ 2007).
Black holes that are emitting below the Eddington luminosity by a factor of $x$
are detectable with ALMA only if their masses exceed $10^5x$ $M_\odot$.
Also, the Eddington time of $\sim 10^8$ yr is 25\% of
the Hubble time at $z=10$, yielding a fair probability for detection of these
systems in the ALMA frequency window.
Finding sources is best done in the continuum (e.g.\ the James Webb
Space Telescope), with ALMA follow-up. Since the molecular lines are optically
thin, they trace the kinematics of all irradiated gas. Multiple CO and H$_2$
lines further allow one to determine the ambient density, temperature and
X-ray flux.

[OI] 63 $\mu$m and [CII] 158 $\mu$m fine-structure lines can be important
coolants of X-ray irradiated gas as well. At a metallicity of $\sim 10^{-2}$,
fine-structure line cooling contributes about half of the total cooling. These
lines could also be observed with ALMA, but FUV or cosmic ray irradiation
boosts them as well (Meijerink et al.\ 2007), diminishing their diagnostic
value for tracing black hole accretion.
In addition, the 149 $\mu$m ($J=1-0$, $v=0$) rotational line of HeH$^+$ is
shown in Figure 2, because it was suggested by Maloney et al.\ (1996)
as a useful XDR tracer. It is found that this line is typically much weaker
than the H$_2$ and CO lines, and HeH$^+$ abundances do not exceed $10^{-9}$.

Fast ($\sim 50$ km/s) shocks can lead to similar CO and H$_2$ emissivities,
albeit over much smaller regions so that beam dilution would be an issue.
Also, shocks should produce CO line profiles with strong non-Gaussian wings.
At redshifts of a few, CO emission with upper levels $J=2-10$ from massive tori
in active galaxies is also accessible to ALMA. This has been studied by
Kawakatu et al.\ (2007) for the case where FUV photons drive the chemistry.
Narayanan et al.\ (2008a) look at simulations of $z\sim 6$ quasars in
$10^{12}-10^{13}$ $M_\odot$ halos and find that the CO is highly excited by
starbursts, peaking at $J=5-8$.
Narayanan et al.\ (2008b) further find that AGN-driven winds may leave
signatures in the CO line emission profile in the form of high velocity peaks
at a few times the circular velocity.
Also, Lintott \& Viti (2006) and Meijerink et al.\ (2007) find an increase
in HCN emission with X-ray flux, but this effect is suppressed at low
metallicity and for densities $\ge 10^5$ cm$^{-3}$.
The X-ray driven lines presented here complement these efforts.
Finally, Figure 3 shows a CO comparison
between an AGN (XDR) and a starburst (PDR = photon dominated region)
model, for the
same impinging flux by energy of 100 erg s$^{-1}$ cm$^{-2}$, typical of a
$10^{44}$ erg s$^{-1}$ Seyfert nucleus or $10^6$ B0 stars within a 200 pc
region. Solar metallicity is assumed, merely because it favors the PDR.
It is obvious that star formation can
never compete with an XDR, for the same illuminating flux by energy, in terms
of the very high $J$ CO line intensities that are produced.

\acknowledgments
The authors are grateful to Paul van der Werf for his comments on H$_3^+$
destruction, to Joe Silk for discussions on CO emission and to the anonymous
referee for his/her very helpful comments.

\clearpage

\begin{figure*}[t]
\centerline{\includegraphics[height=140mm,clip=]{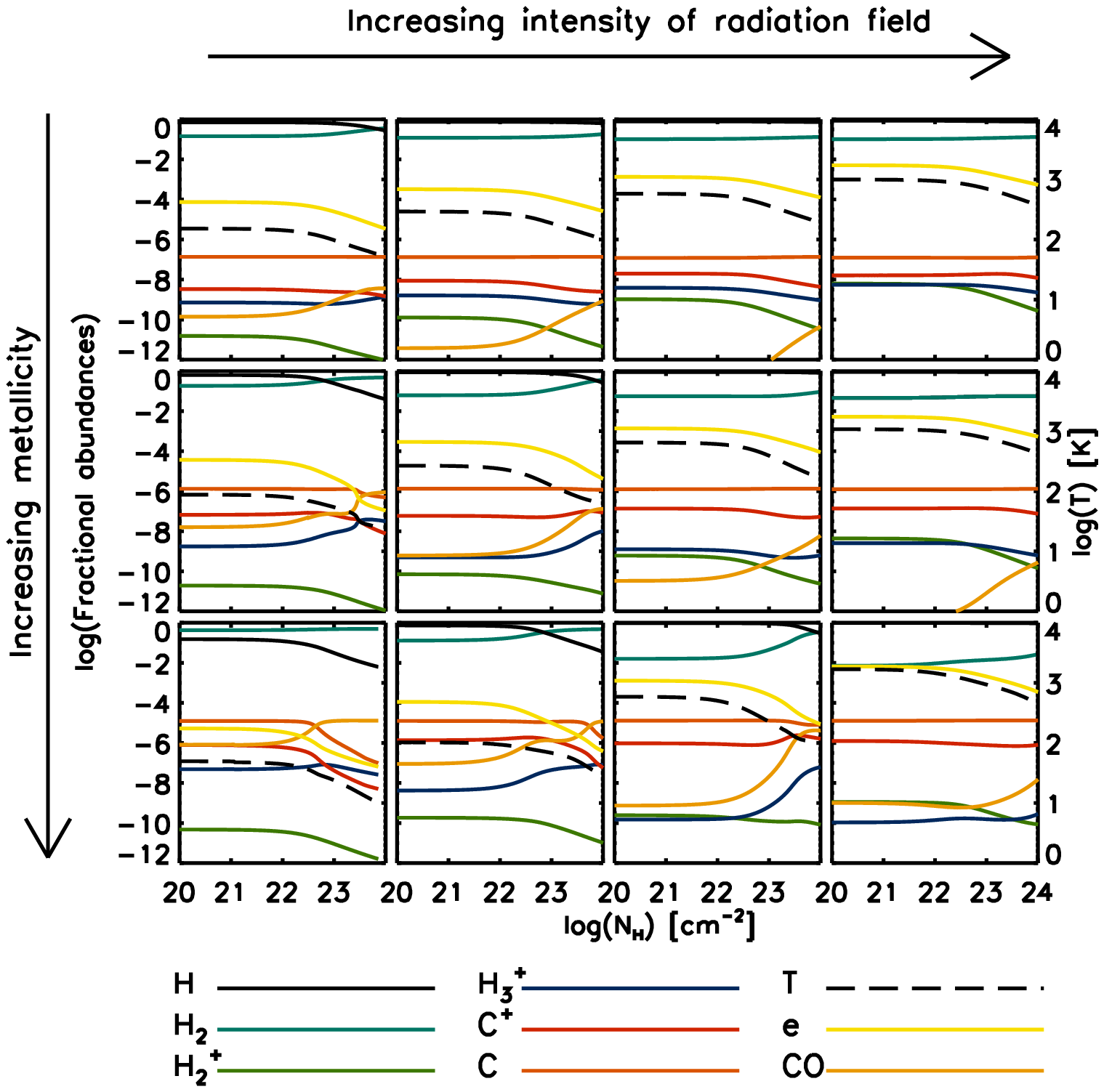}}
\caption{Depth dependence of CO, H$_2$, H$_3^+$, electron abundance,
temperature and other relevant species for various black hole environment
models at a density of $10^5$ cm$^{-3}$.
The impinging X-ray flux takes on values of 0.1, 1, 10 and 100 erg s$^{-1}$
cm$^{-2}$; and the metallicity values of $10^{-3}$, $10^{-2}$ and $10^{-1}$
of solar.}
\label{Chem}
\end{figure*}

\begin{figure*}[t]
\centerline{\includegraphics[height=140mm,clip=]{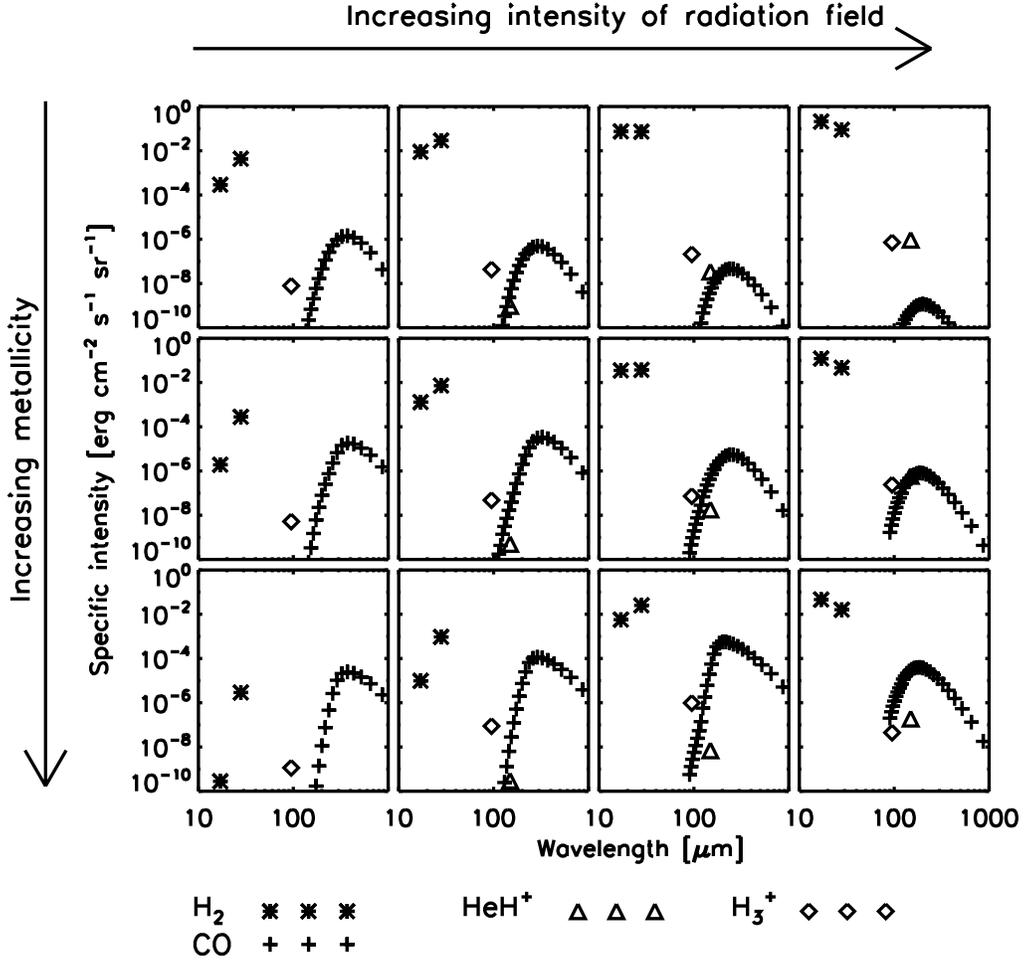}}
\caption{Rest frame spectral line distributions of high J CO, H$_2$ S(0) \&
S(1), H$_3^+$ 95 $\mu$m and HeH$^+$ 149 $\mu$m are shown as functions of
metallicity and X-ray flux for a density of $10^5$ cm$^{-3}$.
The impinging X-ray flux takes on values of 0.1, 1, 10 and 100 erg s$^{-1}$
cm$^{-2}$; and the metallicity values of $10^{-3}$, $10^{-2}$ and $10^{-1}$
of solar.}
\label{SEDs}
\end{figure*}

\begin{figure*}[t]
\centerline{\includegraphics[height=100mm,clip=]{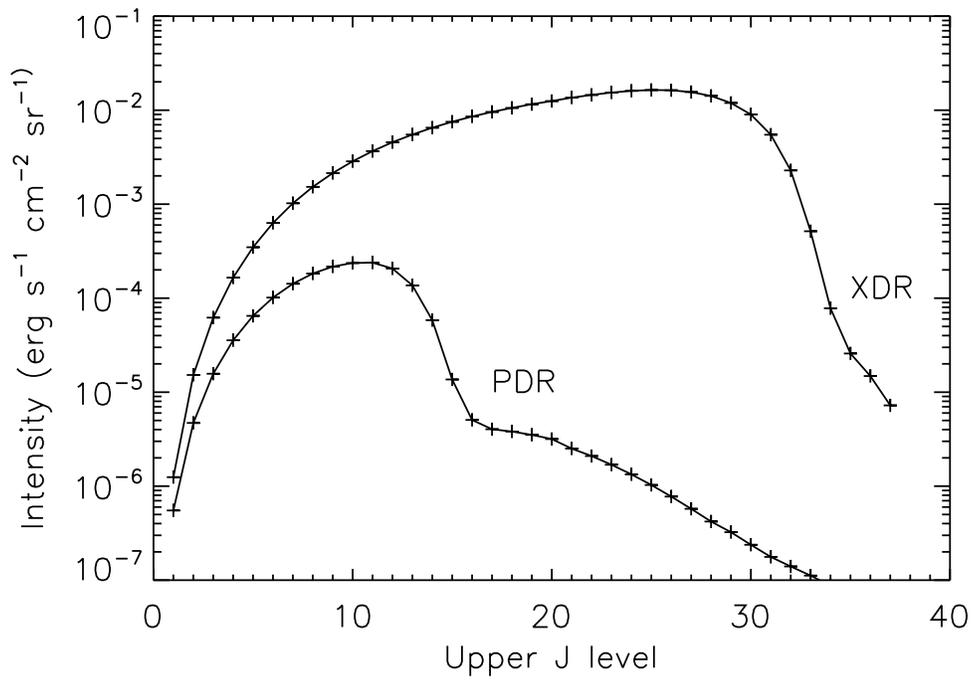}}
\caption{Comparison between an AGN and starburst model, for the same
impinging flux by energy of 100 erg s$^{-1}$ cm$^{-2}$, a density of $10^5$
cm$^{-3}$ and solar metallicity.
The stellar spectrum corresponds to a 30,000 K black body.}
\label{SED}
\end{figure*}

\end{document}